

\input harvmac

\overfullrule=0pt


\def\k{\kappa}
\def\r{\rho}
\def\a{\alpha}

\def\b{\beta}
\def\B{\Beta}

\def\G{\Gamma}
\def\d{\delta}
\def\D{\Delta}
\def\e{\epsilon}

\def\th{\theta}

\def\m{\mu}
\def\n{\nu}

\def\s{\sigma}

\def\no{\noindent}

\def\bs{\bigskip}

\def\qq{\qquad}

\def\bl{\bigl}
\def\br{\bigr}


\def\IR{\relax{\rm I\kern-.18em R}}

\def\gh{$SL(2,\IR)\otimes SO(1,1)^{d-2}/SO(1,1)$}




hep-th/9206xxx \hfill {USC-92/HEP-S1}
\rightline{June 1992}
\bs\bs

\centerline   {\bf CONFORMALLY EXACT RESULTS FOR }
\centerline   {{\bf \gh\ COSET MODELS }
                  {\footnote {$^*$}
                     { e-mail: sfetsos@physics.usc.edu        } }
                   {\footnote{$^\dagger$}
  {Research supported in part by DOE, under Grant
   No. DE-FG03-84ER-40168. } }}

\vskip 1.00 true cm

\centerline {KONSTADINOS SFETSOS}

\bigskip

\centerline {Physics Department}
\centerline {University of Southern California}
\centerline {Los Angeles, CA 90089-0484, USA}


\vskip 1.50 true cm

\centerline{ABSTRACT}

Using the conformal invariance of the \gh\ coset models we calculate
the conformally exact metric and dilaton, to all orders in the
$1/k$ expansion. We consider both vector and axial gauging.
We find that these cosets represent two different space--time geometries:
($2d$ black hole)$\otimes \IR^{d-2}$ for the vector gauging and
($3d$ black string)$\otimes \IR^{d-3}$ for the axial one.
In particular for $d=3$ and for the axial gauging
one obtains the exact metric and dilaton
of the charged black string model introduced by Horne and Horowitz.
If the value of $k$ is finite we find two curvature singularities
which degenerate to one in the semi--classical $k\to \infty$ limit.
We also calculate the reflection and transmission coefficients for the
scattering of a tachyon wave and using the Bogoliubov
transformation we find the Hawking temperature.

\eject

\bs\bs

\vfill
\eject


\newsec{ Introduction }

Models of strings propagating in curved backgrounds has been
studied extensively by means of Conformal Field Theories (CFT),
but most of the effort has been directed
to the case of string compactification where the non--compact part of the
space--time which contains the time coordinate is flat,
i.e described by a trivial CFT, and only the internal part
requires a non--trivial CFT. The particular CFT
used corresponds to a different classical vacuum of the string theory.

In an attempt to formulate solvable models with a single time coordinate
Anti-De-Sitter (ADS) Coset models $G/H=SO(d-1,2)_{-k}/SO(d-1,1)_{-k}$
and their $N=1$ superconformal generalizations
were introduced as exact string theories
\ref\BN{I. Bars and D. Nemechansky, Nucl. Phys. {\bf B348} (1991) 89.}.
The important difference with previous
treatments is that the time direction can be curved as well,
i.e the non--compact part requires a non--trivial CFT.
All single time coordinate models based on simple non--compact groups are
characterized by $G/H$ cosets and the complete list can be found in
\ref\B{I. Bars, ``Curved Space--Time Strings and Black Holes'', in Proc.
of $XX^{th}$ {\it Int. Conf. on Diff. Geometrical Methods in Physics},
Eds. S. Catto and A. Rocha, Vol. 2, p. 695, (World Scientific, 1992).}.
Naturally, by taking non--simple (direct product) groups, one can construct
extensions of these models (for a classification see
\ref\GIN{P. Ginspang and F. Quevedo, ``Strings on Curved Space--Times:
Black Holes, Torsion, and Duality'', LA-UR-92-640.}).
The spectrum of such theories can in principle be found by using
non--compact current algebra techniques
\ref\BL{L. Dixon, J. Lykken and M. Peskin, Nucl. Phys. {\bf B325} (1989) 325.
\semi
I. Bars, Nucl. Phys. {\bf B334} (1990) 125.}.
For all of these cosets the action is written in the form of a gauged WZW
model
\ref\WZW{E. Witten, Nucl. Phys. {\bf B223} (1983) 422. \semi
K. Bardakci, E. Rabinovici and B. Saering , Nucl. Phys. {\bf B301} (1988) 151.
\semi
K. Gawedzki and A. Kupiainen, Nucl. Phys. {\bf B320} (1989) 625.\semi
H. J. Schnitzer, Nucl. Phys. {\bf B324} (1989) 412. \semi
D. Karabali, Q--Han Park, H. J. Schnitzer and Z. Yang,
Phys. Lett. {\bf 216B} (1989) 307.\semi
D. Karabali and H. J. Schnitzer, Nucl. Phys. {\bf B329} (1990) 649.}.
The semi--classical analysis
\ref\WIT{E. Witten, Phys. Rev. {\bf D44} (1991) 314.}
for $k\to \infty$ showed that these are useful models for learning more
about string and particle theories in gravitationally singular spaces.
Higher dimensional models have been subjected since, to the same
semi--classical analysis and various interesting singularities were found
\ref\CRE{M. Crescimanno, Mod. Phys. Lett. {\bf A7} (1992) 489.}
\ref\HOHO{J. H. Horne and G. T. Horowitz, Nucl. Phys. {\bf B368} (1992) 444.}
\ref\BSf{I. Bars and K. Sfetsos, Mod. Phys. Lett. {\bf A7} (1992) 1091.}
\ref\BSs{I. Bars and K. Sfetsos, Phys. Lett. {\bf 277B} (1992) 269.}
\ref\FRA{E. S. Fradkin and V. Ya. Linetsky, Phys. Lett. {\bf 277B} (1992) 73.}
\ref\HORA{P. Horava, Phys. Lett. {\bf 278B} (1992) 101.}
\ref\RAI{E. Raiten, ``Perturbations of a Stringy Black Hole'',
Fermi-Lub 91-338-T.}
\ref\GER{D. Gerson, ``Exact Solutions of Four--Dimensional Black Holes in
String Theory'', TAUP-1937-91.}\GIN.
Cosmological aspects of coset model are discussed in
\ref\VAFA{A. Tseytlin and C. Vafa, Nucl. Phys. {\bf B372} (1992) 443.}
\ref\BSt{I. Bars and K. Sfetsos, ``Global Analysis of New Gravitational
Singularities in String and Particle Theories'',
USC-92/HEP-B1 (hep-th/9205037).}
\ref\KOU{C. Kounnas and D. L\"ust, ``Cosmological String Backgrounds from
Gauged WZW--models ,CERN-TH-6494-92.}.
Heterotic and type--II superstring actions can be constructed \BSs\ in exactly
four space--time dimensions providing useful theories for investigating
the physics of the early Universe in the context of string theory.
It has been shown that the duality properties of the compactified boson
on a circle have their correspondence in this context of string theory
\ref\GIV{A. Giveon, Mod. Phys. Lett. {\bf A6} (1991) 2843.}
\ref\KIR{E. B. Kiritsis, Mod. Phys. Lett. {\bf A6} (1991), 2871.}
\ref\BA{I. Bars, ``String Propagation on Black Holes'', USC-91/HEP-B3.}
\ref\DVV{R. Dijgraaf, E. Verlinde and H. Verlinde,
Nucl. Phys. {\bf B371} (1992) 269.}.
The existence of a {\it discrete} generalized duality is given in \BSf\ while
further dualities based on Killing vectors can be found in
\ref\ROVE{M. Rocek and E. Verlinde, Nucl. Phys. {\bf B373} (1992) 630.}
\ref\ROGI{M. Rocek and A. Giveon, IAS preprint IASSNS-HEP-91/84.}.

The principal method of investigation of the semi--classical geometries
followed ref.\WIT\ that used a Lagrangian method starting from the
gauged WZW action.
In practice, one can use this method to calculate the lowest
fields of the string theory, namely,
the metric $G_{\m\n}$, the antisymmetric tensor $B_{\m\n}$ and,
the dilaton field $\Phi$, to lowest order in the $1/k$ expansion.
The above fields satisfy \WIT\CRE\BSf\BSs\FRA\ the perturbative equations
for conformal invariance
\ref\CALLAN{C. Callan, D. Friedan, E. Martinec and M. Perry,
Nucl. Phys. {\bf B262} (1985) 593.}.
Another drawback of the gauged WZW method is that one obtains
the various fields in only one
patch of the group manifold because of the gauge fixing procedure
\BSf\BSs\FRA. A different gauge choice leads to a metric in a different
coordinate patch which may bear no resemblance to the previous one
(e.g compare \CRE\ to \BSf\ or \FRA).
In \BSt\ a group theoretical method for the global analysis
of any semi--classical geometry, including an explicit solution for the
particle geodesics, was formulated and applied explicitly to some cases.

It is well known that a necessary condition for a critical string theory
requires that the central charge of the matter part exactly compensates
the central charge from the Faddeev--Popov ghosts ($c=-26$ or $c=-15$ when
supersymmetric) so that the trace anomaly vanishes. Most of the CFT
based on coset models require a value for $k$ which is far from being large.
Thus one needs to go beyond the large $k$ limit.
Following a Hamiltonian approach to gauged WZW models the authors of
\ref\BSfo{I. Bars and K. Sfetsos, ``Conformally Exact Metric and Dilaton
in String Theory on Curved Spacetime'', USC-92/HEP-B2 (hep-th/9206006).}
formulated a general
method for computing the conformally exact metric and dilaton,
to all orders in the $1/k$
expansion, for any bosonic, heterotic, or type--II superstring based on a
coset $G/H$ and gave explicit results for the $d=2,3,4$ ADS models. In the
$k\to \infty$ limit these results tend to those one obtains in the
semi--classical approach, to leading order in perturbation theory. In the
special case $d=2$ they were also in agreement with the exact metric and
dilaton obtained in a previous computation \DVV.

In this paper we use the above method to obtain
the conformally exact metric and dilaton for a simple
class of models involving several abelian factors,
i.e \gh\ , which is a $d$--dimensional model.
For $d=3$ the semi--classical aspects of the model were
worked out in \HOHO, for $d=4$ in \RAI\ and for general $d$ in \GIN.
The paper is organized as follows.
In section 2 we review the general method for computing the
conformally exact metric
and dilaton with particular emphasis on the \gh\ coset models.
In section 3 we are dealing with the vector gauge and in
section 4 with the axial one.
In section 5 we consider the scattering of a tachyon wave
in the geometry of the coset manifold and we compute the Hawking
temperature using the Bogoliubov transformation.
Finally we end the paper in section 6 with concluding remarks and discussion.

\bs\bs

\newsec{ The general method and the \gh\ models }

In this section we will briefly review the general method for
computing the conformally exact metric and dilaton fields for any
bosonic $\s$--model based on a coset $G/H$ as it was developed in \BSfo.
Generalizations of the method for the cases where there is superconformal
symmetry can be found in \BSfo.
Let us consider a bosonic string theory for closed strings in $d$ curved
space--time dimensions, based on a $\s$--model CFT with string coordinates
$X^\m,\ \m=0,1\cdots d-1$. We denote the space--time metric and dilaton fields
by $G_{\m\n}(X)$ and $\Phi(X)$ respectively.
We begin with the most general form of the effective action for the
tachyon $T$ in $d$ space--time dimensions

\eqn\action{\eqalign {&S[T]=\int d^d X \sqrt{-G}e^{\Phi}
                     \bl(G^{\m\n}\partial_\m T\partial_\n T-V(T)\br)\cr
                     &V(T)=2T^2+ {\cal O}(T^3)\ ,} }

\no
where $V(T)$ is the tachyon potential whose precise form is not necessary
for the analysis that follows.
{}From the point of view of the CFT the tachyon is completely defined through
the action of the zero modes, $L_0$ and $\bar L_0$,
of the stress tensors for the right and left movers. Therefore \action\
must be equivalent to the following action

\eqn\actionc{S_c[T]=\int d^d X \sqrt{-G} e^{\Phi}
\bl(T (L_0+\bar L_0) T-V(T)\br)\ .}

\no
Comparison of \action\ with \actionc\ determines the form
of $L_0+\bar L_0$ as a differential operator in configuration space

\eqn\lzero{(L_0+\bar L_0)\ T=-{1\over e^{\Phi}\sqrt{-G}}
\partial_\m G^{\m\n}e^{\Phi}\sqrt{-G}\partial_{\n}T\ . }

\no
Using the equivalence between gauged WZW models and current algebra coset
models $G_{-k}/H_{-k}$ we can write $L_0$ in terms of the quadratic
Casimir operators $\D_G$ and $\D_H$ for the group and the subgroup, as follows

\eqn\lzeroco{\eqalign{&L_0T=\bl({\Delta_G\over k-g}
          -{\Delta_H\over k-h}\br)T\cr
          &\Delta_G\equiv Tr(J_G)^2,\qq \Delta_H\equiv Tr(J_H)^2\ ,\cr} }

\no
where $J_G$, $J_H$ are the group and subgroup operators
obeying the appropriate Lie algebras,
and $g$, $h$ are the Coxeter numbers for the group and subgroup respectively.
An expression similar to \lzeroco\ can also be written for $\bar L_0$.
The currents $J_G$, $J_H$, $\bar J_G$, $\bar J_H$ act as first order
differential operators on the group parameter space. Consequently the Casimir
operators $\D_G$, $\D_H$, $\bar \D_G$, $\bar \D_H$ contain single and double
derivatives with respect to all $dimG$ parameters in $G$.
At the tachyon level we require states which are singlets under the gauge
group $H$ (acting simultaneously on left and right movers). Thus we can impose
the following conditions on the tachyon $T$

\eqn\gauge{\eqalign{&(J_H+\bar J_H)\ T=0\ ,\qq{\rm Vector\ gauging}\cr
                    &(J_H-\bar J_H)\ T=0\ ,\qq{\rm Axial\ gauging}\ .\cr}}

\no
The second of the above conditions is appropriate only for the currents
associated with the abelian part of the subgroup.
The number of conditions is $dimH$ and therefore $T$ can only depend on
$d=dim(G/H)$ parameters, $X^\m$ (string coordinates),
which are $H$--invariants. Consequently, using the chain rule, we reduce the
derivatives in \lzeroco\ to only derivatives with respect to the $d$ string
coordinates $X^\m$.
The gauge invariance condition \gauge\ implies that $(\D_H-\bar \D_H)\ T=0$.
Using this and the fact that $\D_G=\bar \D_G$ for any
group \BSfo, we ensure the physical condition for closed bosonic strings
$(L_0-\bar L_0)\ T=0$.
Then using \lzero\ and \lzeroco\ one can deduce uniquely the expression
for the inverse metric $G^{\m\n}$ by comparing the coefficients of the
double derivatives $\partial_\m \partial_\n T$. Comparison of the single
derivative terms $\partial_\m T$ will give a system of $d$ first order
partial differential equations, whose solution determines the
dilaton field $\Phi$.

Let us specialize to the \gh\ coset models. It is convenient to parametrize
the group element of $G=SL(2,\IR)\otimes SO(1,1)^{d-2}$ as follows
\foot{We follow closely the notation of \GIN.}

\eqn\grel{g=\pmatrix{g_0&0&\cdots&0\cr
                     0&g_1&\ldots&0\cr
                     \vdots&\vdots&\ddots&\vdots\cr
                     0&0&\cdots&g_{d-2}\cr},}

\no
where

\eqn\grslr{g_0=\pmatrix{a&u\cr-v&b\cr}, \qq ab+uv=1}

\no
and
\eqn\gru{g_i=\pmatrix{\cosh 2r_i&\sinh 2r_i\cr
\sinh 2r_i&\cosh 2r_i\cr}, \qq i=1,2,\cdots , d-2\ .}

\no
The infinitesimal generators for $SL(2,\IR)$ are

\eqn\gensl{j_0={q_0\over 2}\pmatrix{1&0\cr 0&-1\cr},\quad
j_+={q_0\over 2}\pmatrix{0&1\cr 0&0\cr},\quad
j_-={q_0\over 2}\pmatrix{0&0\cr -1&0\cr}}

\no
and those for the $SO(1,1)$'s

\eqn\genu{j_i={q_i\over 2}\pmatrix{0&1\cr 1&0\cr}, \qq i=1,2,\cdots , d-2\ .}

\no
The coefficients $q_i$ parametrize the embedding of $H=SO(1,1)$ into the
factored $SO(1,1)$'s in $G$ and are normalized to $\sum_{i=0}^{d-2}q_i ^2=1$.
With this normalization the level of $SL(2,\IR)$ is $q_0 ^2 k$
and that of the $SO(1,1)$'s in $G$ is $q_i ^2 k_i$. Therefore the level of
$H=SO(1,1)$ is $q_0 ^2 k+\sum_{i=1}^{d-2} q_i ^2 k_i$.
If we consider the infinitesimal transformations $\d g=g j_a$\ (right)
and $\d g=j_a g$\ (left), where $a=0,\pm$
we find the following expressions for the infinitesimal group generators

\eqn\genv{\eqalign{&J_0={1\over 2}(a \partial_a
-b \partial_b-u \partial_u)\ , \qq \cr
&J_+=a \partial_u-v \partial_b\ ,\cr
&J_-=-u \partial_a\ ,\cr}
\eqalign{&\bar J_0={1\over 2}(b \partial_b
-a \partial_a-u \partial_u)\cr
&\bar J_+=b \partial_u-v \partial_a\cr
&\bar J_-=-u \partial_b\ .\cr}}

\no
In the previous expressions for the generators, $a$, $b$, $u$ were taken
as the independent group parameters, while $v=(1-ab)/u$.
As we shall see they are more
convenient to use in the vector gauge. For the axial gauge
$a$, $u$, $v$ will be used as independent parameters.
In the latter case the generators have the form

\eqn\gena{\eqalign{&J_0={1\over 2}(a \partial_a
-u \partial_u+v \partial_v)\ , \qq\cr
&J_+=a \partial_u\ ,\cr
&J_-=b \partial_v-u \partial_a\ ,\cr}
\eqalign{&\bar J_0={1\over 2}(v \partial_v
-a \partial_a-u \partial_u)\cr
&\bar J_+=b\partial_u-v \partial_a\cr
&\bar J_-=a \partial_v\ .\cr}}

\no
It can easily be shown that the $SL(2,\IR)$ Lie algebra is indeed
obeyed for both the left and the right generators separately and that
any left commutes with any right generator.
For the $SO(1,1)$'s the generators are

\eqn\genuf{\eqalign{&J_i ={1\over 2}q_i \partial_i\ ,\qq\qq\cr
                     &J_H=q_0 J_0+\sum_{i=1}^{d-2}J_i\ ,\cr}
        \eqalign{&\bar J_i=-{1\over 2}q_i \partial_i\cr
             &\bar J_H=q_0 \bar J_0+\sum_{i=1}^{d-2}\bar J_i\ ,\cr}}

\no
where $\partial_i \equiv {\partial\over {\partial r_i}}$.
The central charge for both the right and the left movers is

\eqn\cent{c={3 k\over {k-2}}+(d-2)-1\ .}

\no
Conformal invariance requires that $c=26$.
In what follows we assume that $k>2$ and therefore $d\leq 26$.
If we analytically continue the expressions for the various metrics below,
to the range of parameters $k<2$, $d>26$ we get unphysical metrics.

\bs\bs


\newsec{ The vector gauging }

In this case using \genv\ and \genuf\ the first condition in \gauge\ takes
the following simple form

\eqn\tachv{\partial_u T=0 \quad \Rightarrow \quad T=T(a,b,r_i)\ .}

\no
Then by using \lzero\ and \lzeroco\ we determine the inverse metric
\foot {In what follows we disregard a factor
of ${1\over 2(k-2)}$ in $G^{\m\n}$.
We do the same for the axial gauging as well.}
$G^{\m\n},\ \m,\n=a,b,1,2,\cdots,d-2$ (by comparing the
coefficients of the double derivatives $\partial_\m \partial_ \n T$)

\eqn\invmetv{G^{\m\n}=\pmatrix{\s^2 a^2
&2(ab-1)-\s^2 ab
&-(1-{2/k}){\eta_j \over {1+\r^2}}\ a\cr
2(ab-1)-\s^2 ab
&\s^2 b^2
&(1-{2/k}){\eta_j \over {1+\r^2}}\ b\cr
-(1-{2/k}){\eta_i \over {1+\r^2}}\ a
&(1-{2/k}){\eta_i \over {1+\r^2}}\ b
&(1-{2/k})\bl({\d_{ij}\over \k_i}-{\eta_i\eta_j\over {1+\r^2}}\br)\cr}\ ,}

\no
where $\eta_i\equiv q_i/q_0$, $\k_i\equiv k_i/k$,
$\r^2\equiv \sum_{i=1}^{d-2} \eta_i^2 \k_i$,
$\s^2\equiv {\r^2+2/k\over {1+\r^2}}$,
and we obtain a system of two first order partial differential equations
(by comparing the single derivative terms)
which will determine the dilaton $\Phi$

\eqn\dileqv{\eqalign{&\partial_a\bl(e^{\Phi}\sqrt{-G}G^{ab}\br)+
\partial_b\bl(e^{\Phi}\sqrt{-G}G^{bb}\br)
=2 e^{\Phi}\sqrt{-G}(1+{1\over 2}\s^2) b\cr
&\partial_b\bl(e^{\Phi}\sqrt{-G}G^{ab}\br)+
\partial_a\bl(e^{\Phi}\sqrt{-G}G^{aa}\br)
=2 e^{\Phi}\sqrt{-G}(1+{1\over 2}\s^2) a\ .\cr}}

\no
If we invert the inverse metric we get the following
expression for the line element

\eqn\metv{\eqalign{ds^2=&{1/2\over {1-(1-2/k)ab}}
\bigl[{1\over k}{1\over {ab-1}}(bda+adb)^2-2\ dadb\bigr]\cr
&+{1\over {1-(1-2/k)ab}}\sum_{i=1}^{d-2}(bda-adb)\eta_i\k_i\ dr_i\cr
&+{1\over {1-2/k}}\sum_{i,j=1}^{d-2}\k_i(\d_{ij}
+{\eta_i\eta_j\k_j \over {1-(1-2/k)ab}})\ dr_idr_j\ . \cr}}

\no
The solution to the system of differential equations gives
the following result for the dilaton

\eqn\dilv{C\ e^{\Phi}=(1-ab)\sqrt{1+{2\over k}{ab\over {1-ab}}}\ ,}

\no
where $C$ is the constant of integration.
We have thus found the exact metric and dilaton
for the vector gauging of the \gh\ model.
\foot {The combination $e^{\Phi} \sqrt{-G}$ is $k$--independent,
as it was conjectured in previous work \BSf\BSs\BSfo.
The same is true for the case of the Axial gauging.}
In the $k\to \infty$ limit eq.\metv\ agrees with the semi--classical
expression found in \GIN.
The dilaton in \dilv\ is independent of
the $r_i$ coordinates and it is the same as
the exact dilaton found in \DVV\ for the $2d$ black hole.
This fact gives the hint that the two models are very
closely related. Indeed as in \GIN\ we can show that there is a
coordinate transformation which diagonalizes the metric.
In the region where $ab>1$ we can make the following transformations

\eqn\transf{\eqalign{&a=\cosh R\ e^{X_0+mX_{d-2}},
\qq b=\cosh R\ e^{-(X_0+mX_{d-2})}\cr
&r_i=\sqrt{1-2/k}\ N_{ij}X_j\ ,\cr}}

\no
with

\eqn\chco{N_{ij}=\cases{
-{\textstyle\rho_j\over \textstyle\rho_i\sqrt {\kappa_i}}&$i=j+1$\cr
\noalign{\vskip2pt}
{\textstyle\sqrt{\kappa_{j+1}}\,\eta_i\, \eta_{j+1}
\over\textstyle \rho_{j+1}\rho_j}&$i\leq j\not= d-2$\cr
\noalign{\vskip2pt}
{\textstyle\eta_i\over \textstyle\rho_j(\rho_j^2+1)^{1/2}}&$i\leq j=d-2$\cr
\noalign{\vskip2pt}
\qq 0&otherwise \cr}\ , \qq m=-\sqrt{1-2/k}\ {\r\over (1+\r^2)^{1\over 2}}\ ,}

\no
where

\eqn\para{\r_i^2=\sum_{j=1}^i \k_j \eta_j ^2\ ,\qq {\rm and}\qq
                 \r_{d-2}\equiv \r\ .}

\no
The matrix elements $N_{ij}$ satisfy the relations

\eqn\ide{\eqalign{
&\sum_{l=1}^{d-2}\k_l\ N_{li}\ N_{lj}=\d_{ij} \qq i,j\not=d-2\cr
&\sum_{l=1}^{d-2}\k_l\ N_{l,d-2} ^2={1\over {1+\r^2}}\cr
&\sum_{l=1}^{d-2}\k_l\ \eta_l\ N_{li}=0 \qq i\not=d-2\cr
&\sum_{l=1}^{d-2}\k_l\ \eta_l\ N_{l,d-2}
={\r\over (1+\r^2)^{1\over 2}}\ .\cr}}

\no
In these new coordinates the metric takes the form

\eqn\metvc{ds^2=dR^2-{1\over {\tanh^2 R-2/k}}\ dX_0^2
+\sum_{i=1}^{d-2}dX_i ^2\ .}

\no
The first two terms in \metvc\ are the exact metric found
in \DVV\ for the $SL(2,\IR)/SO(1,1)$ $2d$ black hole.
Although the embedding of $H=SO(1,1)$ in $G$ was general the resulting
geometry coincides with the case $\eta_i =0$, i.e $H=SO(1,1)$ embedded
only in $SL(2,\IR)$. This is as expected because for the vector
gauging $\d r_i=0$. Therefore we have proved that the \gh\ model
for the vector gauging, is equivalent to the
$(2d\ {\rm black\ hole})\otimes  \IR^{d-2}$ model for any $k$.
In the semi--classical limit $k\to \infty$ this was proved in \GIN.
For the regions $ab<0$ and $0<ab<1$ we can find the conformally
exact metric by analytically continue $R\to R+i\pi/2$
and $R\to it$ respectively.

\bs\bs


\newsec{ The axial gauging }

The most interesting case is that of the axial gauging.
Then using \gena\ and \genuf\ the second condition in \gauge\ becomes

\eqn\tacha{(a\partial_a+\sum_{i=1}^{d-2}\eta_i\partial_i)\ T=0
\quad \Rightarrow \quad T=T(u,v,x_i=r_i-\eta_i\ {\rm ln}a)\ .}

\no
Proceeding as in the previous section for the vector
gauging we find for the inverse metric the following expression
\foot {See remarks in footnote 2.}

\eqn\invmeta{G^{\m\n}=\pmatrix{\s^2 u^2
&2(uv-1)-\s^2 uv
&-u \eta_j\cr
2(uv-1)-\s^2 uv
&\s^2 v^2
&-v \eta_j\cr
-u \eta_i
&-v \eta_i
&(1-2/k){\d_{ij}\over \k_i}+\eta_i\eta_j \cr}\ .}

\no
The differential equations which determine the dilaton
are similar to \dileqv\ above with $(a,b)\to (u,v)$.
If we invert the inverse metric and solve the system of
differential equations we find for the line element

\eqn\metxa{\eqalign{ds^2=&{1\over{[(1-2/k)(uv-1)
-\r^2-2/k][(1-2/k)(uv-1)-\r^2]}}\biggl(
-{1-2/k \over 2k} (v^2 du^2 +u^2 dv^2)\cr
&+\bigl[(1-2/k)\bl((1-1/k +\r^2)(uv-1)-1/k\br)-\r^2 (1+\r^2)\bigr]\ dudv\biggl)
\cr
&+{1\over {(1-2/k)(uv-1)-\r^2}}
\sum_{i=1}^{d-2}(vdu+udv)\eta_i \k_i\ dx_i\cr
&+{1\over {1-2/k}}\sum_{i,j=1}^{d-2}\k_i(\d_{ij}
+{\eta_i \eta_j \k_j
\over {(1-2/k)(uv-1)-\r^2}})\ dx_idx_j \cr}}

\no
and for the dilaton

\eqn\dila{C' e^{\Phi}=(1-uv)\sqrt{[1+\r^2
+(\r^2+2/k){uv\over {1-uv}}][1+\r^2-2/k+\r^2{uv\over {1-uv}}]}\ ,}

\no
where $C'$ is the constant of integration.
Thus, we have obtained the exact expressions for the metric
and the dilaton of the \gh\ model in the
axial gauging generalizing the previous semi--classical
results of \HOHO\ for $d=3$, \RAI\ for $d=4$ and \GIN\ for any $d$.
\foot {See remarks in footnote 3.}
Our expression for the metric \metxa\ is not yet
ready to be compared with the corresponding semi--classical
expression in \GIN. To do so we must specialize to the ``gauge''

\eqn\ga{b=\pm a \quad \Rightarrow \quad x_i
=r_i-{1\over 2}\eta_i\ \ln |1-uv|\ .}

\no
Under this change the dilaton is unaffected (still given by \dila) whereas
the metric takes the following form

\eqn\meta{\eqalign{ds^2=&{1\over {(1-2/k)(uv-1)-\r^2-2/k}}\bigl[-
{1\over 4}{\r^2+2/k\over {uv-1}}(vdu+udv)^2
+(1+\r^2)\ dudv\bigr]\cr
&+{1\over {1-2/k}}\sum_{i,j=1}^{d-2}\k_i(\d_{ij}
+{\eta_i\eta_j\k_j\over {(1-2/k)(uv-1)-\r^2}})\ dr_idr_j\ ,\cr}}

\no
which in the $k\to \infty$ limit is in agreement with the
semi--classical expression found in \GIN.
The metric \meta\ is singular at $uv=1$ where the ``gauge choice''
breaks down. This is obviously a coordinate singularity
because \metxa\ is manifestly non--singular at $uv=1$.
We can easily show that the same change
of coordinates \transf \chco \para,
made for the vector case, also diagonalizes the  metric \meta\ but now
with $m=0$. After rescaling $X_0 \to X_0/(1+\r^2)^{1\over 2}$ and
$X_{d-2} \to X_{d-2} ({1+\r^2 \over {1-2/k}})^{1\over 2}$
the metric in the region where $uv>1$ takes the following form

\eqn\metg{ds^2=dR^2
-{dX_0^2 \over {(1+\r^2)\tanh ^2 R-\r^2-2/k}}
+{\tanh ^2 R\over {(\r^2+1-2/k)\tanh ^2 R-\r^2}}\ dX_{d-2}^2
+\sum_{i=1}^{d-3} dX_i ^2\ .}

\no
In the regions $0<uv<1$ and $uv<0$ we obtain the
metric by making the same
analytic continuations of $R$ as in the vector case.
To obtain further insight let us concentrate on the first three
terms of the metric. Inspired by the work in ref. \HOHO\ we make the
following change of variables
in the $uv>1$ region\ (a similar change can be made in the other two regions).

\eqn\var{\cosh ^2 R={r_+ -r \over {r_+ -r_-}}\ ,}

\no
where
\eqn\varr{r_+=M\equiv \sqrt{2/k'}\ (\r^2+1)\ e^a\ ,
\qq r_-=Q^2/M \equiv \sqrt{2/k'}\ (\r^2+2/k)\ e^a\ .}

\no
The constant $a$ is related to $C'$ in \dila\ and $k'=k-2$
is the renormalized value for the central extension $k$.
After a few rescalings of the variables, the $3d$ non--trivial part
of the metric \metg, and the dilaton \dila\ take the following forms


\eqn\trdm{ds_{3d}^2=-(1-{r_+ \over r})\ dt^2
+(1-{r_- -r_q \over {r-r_q}})\ dx^2 +
{k'\over {8r^2}}(1-{r_+ \over r})^{-1}(1-{r_- \over r})^{-1}\ dr^2}

\no
and

\eqn\trddil{\Phi={1\over 2}\ln \bl(r(r-r_q)\br)+{1\over 2}\ln k'\ ,}

\no
where $r_q \equiv 2/k \sqrt{2/k'}\ e^a$.
Notice that $r_q \to 0$ when $k\to \infty$.
The scalar curvature for the metric \trdm\ can also be calculated


\eqn\curv{\eqalign{R=&{4\over {k' \bl(r(r-r_q)\br)^2}}
\bigl\{2(r_+ +r_- -r_q)\ r^3-\bl(7r_+ r_- -r_q (r_- -r_q)\br)\ r^2 \cr
&+r_q r_+ (7r_- +r_q)\ r-3r_q ^2 r_+ r_- \bigr\}\ .\cr}}

\no
The above expressions are the conformally exact
metric, dilaton and scalar curvature
of the $3d$ model which was analyzed by Horn and Horowitz \HOHO\ in the
semi--classical limit. These authors showed that, in the $k\to \infty$ limit,
the metric describes a black string
with mass $M$ and charge $Q$ the same as the quantities defined
in \varr\ in the large $k$ limit. It can be seen by inspecting \curv\ that
now the exact metric \trdm\ has two true curvature singularities
at $r=0$, $r=r_q$ which degenerate to only one singularity, at $r=0$,
in the $k\to \infty$ limit.
There are also two coordinate singularities at $r=r_+$ and $r=r_-$ whose
interpretation will be given in the next section.
It is interesting to take the $\r^2 \to 0$ limit.
In this case one should recover
the $2d\ ({\rm black\ hole}) \otimes \IR$ model since the subgroup $H=SO(1,1)$
is totally embedded in the $SL(2,\IR)$ factor in $G$. It can be checked that
this is indeed the case. In particular the scalar curvature \curv\ becomes
$R=8/k' (r-3/k)/ r^2$ which has only one singularity at $r=0$,
in agreement with \BSfo.

Therefore we have proven that the \gh\ coset model for the
axial gauging is equivalent
to the $(3d\ {\rm black\ string})\otimes \IR^{d-3}$ model for any $k$.
In the semi--classical limit this was proved in \GIN.

Reversing the sign of $M$ is equivalent to reversing the
signs of $r$, $r_-$ and $r_q$. Therefore we restrict ourselves to $M>0$.
As in \HOHO\ we can distinguish three different cases

\medskip
\no
$(i)\ {\rm The\ black\ string\ with}\ 0<Q<M\ (0<r_- < r_+)$

\medskip
This is the generic case. As in the semi--classical $k\to \infty$ limit,
the coordinate singularities at $r=r_+$ and $r=r_-$ can be interpreted
as an event and an inner horizon respectively.

To see the effect the finite value of $k$ has on the structure of the manifold
we will consider the geodesic equations. They have the following form

\eqn\geo{{k'\over 8}\dot r^2=E^2 r(r-r_-)
-P^2(r-r_q)(r-r_+)+\a (r-r_+)(r-r_-)\ ,}

\no
where $E$, $P$ are two conserved quantities,
determined by the initial conditions,
associated with the two killing vectors in the $t$ and $x$ directions
and $\a=0\ (-1)$ for null\ (time--like) trajectories.
For large $r$, the right hand side is non--negative if $E^2-P^2 +\a \geq 0$.
Let us first consider the time--like trajectories.
We can check that the right hand side of \geo\ becomes
negative at $r=r_q >0$. Therefore trajectories reach a minimum value
for $r$ and they never reach either singularity.
The same is true in the semi--classical
limit $k\to \infty$, where no trajectory can hit
the $r=0$ singularity. Now we turn to the case of null trajectories.
In the $k\to \infty$ limit we can prove using \geo\ that
for $E^2-P^2>0$

\eqn\min{r_{\rm min}=\cases{0&\qq {\rm if}
\quad $E^2 r_- -P^2 r_+ \leq 0$\cr
{{E^2} r_- -P^2 r_+\over {E^2-P^2}}&\qq {\rm otherwise}\cr}}

\no
and for $E^2-P^2=0$

\eqn\minm{r_{\rm min}=0\ ,\qq {\rm if}\quad E^2-P^2=0\ .}

\no
Therefore under certain initial conditions null trajectories can
arrive to the singularity at $r=0$.
Qualitatively the behavior is similar to the case of
the Reissner--N\"ordstrom metric of Einstein's general relativity.
However when $k$ is finite the situation changes drastically.
In contrast with the $k\to \infty$ case null trajectories
can never hit the $r=r_q$ singularity.
Instead they reach a minimum value which can be found using \geo

\eqn\rrmin{r_{\rm min}=\cases{r_0/2+\sqrt{r_0 ^2 /4 +P^2/(E^2-P^2)r_q r_+}
&\qq $E^2-P^2>0$\cr
\noalign{\vskip2pt}
2/k\ r_+&\qq $E^2-P^2=0$\ .\cr}}

\no
with

\eqn\ro{r_0={E^2 r_- -P^2(r_+ +r_q)\over {E^2-P^2}}\ .}

\no
In all cases the turning point lies inside the inner horizon.
In the region inside the two singularities no time--like or null
trajectory is allowed because in such case the right hand side of \geo\
is manifestly negative. This is related to the fact that in this region
all variables ($t$, $x$ and $r$) become space--like as one can see by
inspecting \trdm.
Finally, we consider trajectories in the region where $r$ takes
negative values. It can easily be seen that, null trajectories reach the
singularity at $r=0$ only if $P^2=0$.
In contrast if $k\to \infty$ this is possible for $E^2 r_- -P^2 r_+\geq 0$.
In either case, time--like trajectories never hit the singularity.

Now we consider some thermodynamic properties of the black string.
In general, one can deduce the Hawking temperature
associated with the event horizon by considering the
metric in the Euclidean regime
$t\to i\th$, in the neighborhood of the event horizon.
Then if we introduce the parametrization

\eqn\temc{r=r_+ (1+\b^2 z^2)\ ,
\qq \b^2={2\over k'}\bl(1-{r_- \over r+}\br)\ .}

\no
the metric \trdm\ close to the horizon $r=r_+\ (z=0)$ can be written as

\eqn\meteucl{ds_E ^2 \sim dz^2+\b^2 z^2\ d\th^2
+{r_+ -r_-\over {r_+ -r_q}}\ dx^2\ .}

\no
The horizon represents a conical singularity
of the solutions of the Euclideanized equations
which can be removed if the imaginary time $\th$ is
taken to be periodic with
$(\rm period)=2\pi/ \b$. The
temperature is identified with the inverse period
\ref\HAWK{J. B. Hartle and S. W. Hawking Phys. Rev. {\bf D13} (1976) 2188.
\semi
S. W. Hawking, Phys. Rev. {\bf D18} (1978) 1747.}.
Therefore the temperature of the black string is

\eqn\temp{T={1\over \pi}\sqrt{{1\over {2k'}}
\bl(1-{r_- \over r_+}\br)}\ .}

\no
This is of the same form (except for the replacement $k\to k'$)
as the expression for the temperature found in \HOHO.
The statistical description of the Hawking radiation
is inappropriate when the back reaction of the emitted radiation starts
to become important
\ref\PRE{J. Prescill, P. Schwarz, A. Shapere, S. Trivedi and F. Wilczek,
Mod. Phys Lett. {\bf A6} (1991) 2353 \semi
C. Holzhey and F. Wilczek, IAS preprint, IASSNS-HEP-91/71.}.
For the black string this happens in the extremal limit (see below).
\foot {One condition which must be satisfied for the statistical description
to be valid is ${\partial T\over \partial M}{\big |}_Q << 1$ \PRE.
This condition is catastrophically violated in the extremal limit.}
We will reevaluate the Hawking temperature in the
next section using the Bogoliubov transformation.

\medskip
\no
$(ii)\ {\rm The\ extremal\ limit}\ Q=M\ (r_-=r_+)$

\medskip
In the limit where $q_0\rightarrow 0\ (\r^2 \rightarrow \infty)$
or equivalently $Q\rightarrow M\ (r_-\rightarrow r_+)$
the embedding of $H=SO(1,1)$ inside $SL(2,\IR)$ is zero
and therefore we expect that the metric \trdm\ reduces
to the metric appropriate for the Anti-de-Sitter
space manifold of $SL(2,\IR)\sim SO(2,1)$.
Indeed if we change variables to

\eqn\chva{y={8({r\over r_+}-1)\over {k'(1-{r_-\over r_+})^{1/2}}}\ ,
\quad \hat t=\br(1-{r_- \over r_+}\br)^{1/4} t\ , \quad
\hat x={(1-{r_-\over r_+})^{1/4} \over (1-{r_-\over r_+})^{1/2}} x}

\no
and take the double scaling limit $r\to r_+$ and
$r_-\to r_+$ the metric \trdm\
can be written as

\eqn\anti{ds_{ads} ^2={k'\over 8}\bl(y(-d\hat t^2
+d\hat x^2)+{1\over y^2}\ dy^2 \br)\ .}

\no
The fact that it describes an Anti-de-Sitter space can be seen
by noticing the boost invariance along the string
and by calculating the Ricci tensor.
The latter reads $R_{\m\n}=-{4\over k'} g_{\m\n}$. One notices that
the geometry is non--singular, but there is still a horizon at $y=0$.
The metric \anti\ is exactly the same as the one found in \HOHO\ in the
semi--classical limit.
\foot{One can check that in the extremal limit the scalar
curvature \curv\ reduces to $R=-12/k'$ which is compatible with the result
for the Ricci tensor above.}
This was expected by the authors of
\ref\SANTA{J. H. Horne, G. T. Horowitz and A. R. Steif,
Phys. Rev. Lett. {\bf 68} (1991) 568.}
on the basis that after some appropriate transformations the metric \anti\
describes a plane--front wave of the same type several authors
\ref\SANT{R. Guven, Phys. Lett. {\bf 191B} (1987) 275.\semi
D. Amati and C. Klimcik, Phys. Pett. {\bf 219B} (1989) 443.\semi
G. T. Horowitz and A. R. Steif, Phys. Rev. Lett. {\bf 64} (1990) 260.}
proved that it solves the $\s$--model perturbative equations to all
orders in the string coupling $(1/k)$.
As we have seen this follows trivially in the Hamiltonian formalism.

\medskip
\no
$(iii)\ {\rm The\ solution\ for\ M<Q}\ (r_+ < r_-)$

\medskip
It can be seen from the definitions \varr\ that the conformal field theory
construction we have followed so far allows only solutions with $Q<M$.
However, as in \HOHO, if we gauge a different subgroup
of $SL(2,\IR)$, namely that generated
by $(j_+ +j_-)$ in \gensl\ we get solutions with $M<Q$. As in \HOHO\
we can obtain
those solutions by setting $\tilde r^2=r-Q^2/M$ in \trdm. Then the metric
reads (using a notation with $M$ and $Q$ this time)

\eqn\trdma{\tilde ds_{3d}^2=-{Q^2-M^2+M\tilde r^2 \over {Q^2
+M\tilde r^2}}\ dt^2 +{M\tilde r^2 \over {Q^2-Mr_q+M\tilde r^2}}\ dx^2
+{k'\over 2}{M\over {Q^2-M^2+M\tilde r^2}}\ d\tilde r^2\ .}

\no
This metric, for $0<\tilde r < \infty$ has no horizons and no
curvature singularities. However, it does have a conical singularity
which can be removed by identifying $x$ with period

$$\pi\sqrt{2k'{Q^2-Mr_q\over {Q^2-M^2}}}\ .$$

\no
As in the semi--classical case \HOHO, this changes
the structure of the space-time at infinity from
$\IR ^3 \to \IR ^2 \times S^1$. In fact, if we take the limit
$M\to 0$, $Q\to 0$ keeping $Q^2/M=fixed$ the
metric \trdma\ reduces to the sum of $-dt^2$ and the conformally exact metric
for the Euclidean $2d$ black hole \WIT\DVV.

\bs\bs

\newsec{Scattering off the black string.}

In this section will describe the scattering of a tachyon wave
off the black string, in the generic case $0<Q<M$,
by solving \lzero\ for the $3d$ metric \trdm\ and the dilaton \trddil.
Because of the independence of this equation on the variables
$t$ and $x$ we look for solutions with the following form

\eqn\repl{\eqalign{&T(t,x,r)=e^{-iEt} e^{-iNx} T(r)\cr
&E=4\bl(2k(\r^2+1)\br)^{-1/2} \m\ , \quad N=4(2k \r^2)^{-1/2} \n\ ,\cr}}

\no
where $\m, \n\in R$ and the various factors in the expressions for $E$, $N$
were introduced for later convenience. We change variables to

\eqn\allagi{\eqalign{&z={r_+ -r\over {r_+ -r_-}}\cr
&T=z^{c-1\over 2}(1-z)^{a+b-c\over 2} \Psi\ ,\cr}}

\no
where the constants $a,b,c$ are defined as

\eqn\statheres{\eqalign{&a=j+1+i(\e |\m|-\e'|\n|)\cr
&b=-j+i(\e |\m|-\e'|\n|)\cr
&c=1+2 i\e |\m|\cr
&\e,\e'=\pm\ ,\cr}}

\no
provided the eigenvalue of $L_0$, takes the appropriate form for a coset
model

\eqn\idiotimi{\eqalign{L_0&=-{j(j+1)\over {k-2}}-{1\over {k}}
\bl({\m^2\over {\r^2+1}}-{\n^2\over \r^2}\br)\cr
&=-{1\over k'}\bl(j(j+1)
+(r_+ -r_-)({\m^2 \over r_+} -{\n^2 \over r_- -r_q})\br)\cr}}

\no
and $j$ takes values in a representation of $SL(2,\IR)$
\ref\WYB{B. G. Wybourn, Classical Groups for Physicists
(John Wiley \& sons, 1974).}.
Then $\Psi$ satisfies the standard hypergeometric equation

\eqn\hyper{\bl(z(1-z){d^2\over dz^2} +(c-(1+a+b)z){d\over dz}-ab\br)\Psi=0\ ,}

\no
with solution for $|z|\leq 1$

\eqn\lysis{\Psi(z)=c_1 F(a,b,c;z)+c_2 z^{1-c} F(a+1-c,b+1-c,2-c;z)\ ,}

\no
where $c_1,c_2$ are two arbitrary constants, which can be determined by
imposing the appropriate boundary conditions.
We want to describe the scattering of the tachyon off the black
string geometry. As we shall see, in this case, $j$ should belong to the
principal series representation
of $SL(2,\IR)$ i.e $j=i\s-{1\over 2}, \s \in \IR$.
Let us first consider a solution which,
in the asymptotically flat region $r\to \infty$,
reduces to the sum of two waves, one ingoing and the other outgoing,
and represents a wave which disappears into the event
horizon for $r\to r_+$. We call this type of solutions $T_{out}$ for
reasons which will become apparent.
The appropriate choice for the various constants is
$c_2=0$, $\e=-1$ and $\e'=1$. Then one can check that indeed the solution
has the right asymptotic behavior

\eqn\bound{T_{out}\sim (-1/z)^{i |\m|}\ ,\qq z\to 0^- \ (r\to (r_+)^+)}

\no
and
\eqn\boundi{T_{out}\sim {\G(c)\G(b-a)
\over {\G(b)\G(c-a)}}\ (-z)^{-i\s-1/2}+
{\G(c)\G(a-b)\over {\G(a)\G(c-b)}}(-z)^{i\s-1/2}\ ,
\quad z\to -\infty\ (r\to +\infty)\ .}

\no
where the first term in \boundi\ represents an ingoing wave and the second
an outgoing one. The expressions for the reflection and transmission
amplitudes are

\eqn\reftra{\eqalign{&R_+
={\cosh \pi(\s-|\m|-|\n|)\ \cosh \pi(\s-|\m|+|\n|)
\over {\cosh \pi(\s+|\m|+|\n|)\ \cosh \pi(\s+|\m|-|\n|)}}\cr
&T_+={\sinh 2\pi\s\ \sinh 2\pi |\m| \over
{\cosh \pi(\s+|\m|+|\n|)\ \cosh \pi(\s+|\m|-|\n|)}}\ .}}

\no
We see that part of the wave gets reflected in the event horizon.
The other part will enter the event horizon and will be absorbed
by the black string. The same is true in the
case of the $2d$ black hole \DVV\ for which the results
follow from our formulas if we set $\n=0$.
A similar analysis for scattering in the naked
singularity region, where $r<0$,
gives for the reflection and transmission amplitudes the following results

\eqn\reftran{\eqalign{&R_-
={\cosh \pi(\s-|\m|-|\n|)\ \cosh \pi(\s+|\m|-|\n|)
\over {\cosh \pi(\s+|\m|+|\n|)\ \cosh \pi(\s-|\m|+|\n|)}}\cr
&T_-={\sinh 2\pi \s\ \sinh 2 \pi |\n| \over
{\cosh \pi(\s+|\m|+|\n|)\ \cosh \pi(\s-|\m|+|\n|)}}\ .\cr}}

\no
We see that the naked singularity is not a perfect
reflector as in the $2d$ case \DVV.

Next we construct a solution, which we call $T_{in}$, by imposing
different boundary conditions. Namely, we demand a solution which,
close to the event horizon $r\to r_+$ behaves
as the sum of an ingoing with an outgoing wave, and reduces
to an ingoing wave for $r\to \infty$.
Then the appropriate choice for the constants is $\e=-1$, $\e'=1$ and

\eqn\cc{{c_1\over c_2}=-{\G(2-c)\G(a)\G(c-b) \over \G(c)\G(a+1-c)\G(1-b)}\ .}

\no
The asymptotic behavior now is

\eqn\boun{T_{in}\sim c_1(-1/z)^{i |\m|}+c_2(-1/z)^{-i |\m|}\ ,
\qq z\to 0^- \ (r\to (r_+)^+)}

\no
and
\eqn\bouni{T_{in}\sim \bl(c_1 {\G(c)\G(b-a) \over \G(b)\G(c-a)}
+c_2 {\G(2-c)\G(b-a) \over \G(b+1-c)\G(1-a)}\br) (-z)^{-i\s-1/2}\ ,
\quad z\to -\infty\ (r\to +\infty)\ .}

\no
One can easily see that

\eqn\inout{T_{in}(z;\s,|\m|,|\n|)=c_1 T_{out}(z;\s,|\m|,|\n|)
+c_2 T_{out}^*(x;-\s,|\m|,-|\n|)\ .}

\no
The states $T_{in}$ and $T_{out}$ form two different bases in terms
of which any state can be expanded. Consequently there are two distinct
Fock spaces corresponding to two different vacua.
One can follow a standard procedure (see for instance
\ref\BIR{Quantum Fields in Curved Space, N. D. Birrell and P. C. W. Davies,
Cambridge University Press.})
to show that the expectation value of the occupation number operator $N_{out}$
for the $\{T_{out}\}$ basis in the vacuum of the $\{T_{in}\}$ basis is

\eqn\occ{\eqalign{{}_{in}\langle 0|N_{out}|0\rangle_{in}
&={1\over {|c_1|^2\over |c_2|^2}-1}\cr
&=\biggl({\cosh \pi(\s+|\m|-|\n|)\ \cosh \pi(\s+|\m|+|\n|) \over
\cosh \pi(\s-|\m|-|\n|)\ \cosh \pi(\s-|\m|+|\n|)}-1\biggr)^{-1}\ .\cr}}

\no
This is not zero indicating the fact that the two bases are inequivalent.
We can define the Hawking temperature $T$ by rewriting \occ\ in the following
form

\eqn\occl{{}_{in}\langle 0|N_{out}|0\rangle_{in}\equiv
{1\over e^{M\over T}-1}\ ,}

\no
where $E$, defined in \repl, is the eigenvalue of the time--like
vector $i\partial_t$. One can easily show that, when $\s\to \infty$,
$T$ tends to the same temperature defined in \temp\ above. For
$\s$ small the corresponding expression is different and depends explicitly
on the value of $\s$. We see that for
the ``out'' observers the ``in'' vacuum is full of particles in a
heat bath at temperature $T$.


\newsec{ Discussion and concluding remarks}

By making use of the conformal properties of
the \gh\ coset model we proved, for any $k$, that
it describes geometries equivalent to
the $(2d\ {\rm black\ hole})\otimes \IR^{d-2}$
model for the vector gauging, and to
the $(3d\ {\rm black\ string})\otimes \IR^{d-3}$ model for
the axial one. We gave the conformally
exact expressions for the metric the dilaton and the scalar curvature.
We have seen for the $3d$ case that the finite value
of $k$ has some important consequences.
One of them is the appearance of a second curvature singularity not present
in the semi--classical $k\to \infty$ limit.
Finally we calculated the transmission and reflection
coefficients for the scattering of the tachyon off the black string and
using the Bogoliubov transformation we found the Hawking temperature.
According to \metg\ same conclusions follow for $d\geq 4$.

\bs\bs


\newsec{Acknowledgements}

I would like to thank I. Bars for the numerous stimulating discussions we had.

\listrefs
\end